\documentclass[10pt,twocolumn,letterpaper]{article}

\usepackage[pagenumbers]{iccv} 
\usepackage{adjustbox}
\usepackage{multirow}
\definecolor{iccvblue}{rgb}{0.21,0.49,0.74}
\usepackage[pagebackref,breaklinks,colorlinks,allcolors=iccvblue]{hyperref}

\def\paperID{*****} 
\def\confName{ICCV}
\def\confYear{2025}

\title{\ SHIELD:\ A Secure and Highly Enhanced Integrated Learning for Robust Deepfake Detection against Adversarial Attacks}

\author{Kutub Uddin \quad  Awais Khan \quad  Muhammad Umar Farooq \quad  Khalid Malik \\
College of Innovation \& Technology,\\ University of Michigan-Flint\\
MI, 48502, USA\\
{\tt\small  kutub@umich.edu, mawais@umich.edu, mufarooq@umich.edu, drmalik@umich.edu}
}

\begin{document}
\maketitle
\begin{abstract}
Audio plays a crucial role in applications like speaker verification, voice-enabled smart devices, and audio conferencing. However, audio manipulations, such as deepfakes, pose significant risks by enabling the spread of misinformation. Our empirical analysis reveals that existing methods for detecting deepfake audio are often vulnerable to anti-forensic (AF) attacks, particularly those attacked using generative adversarial networks. In this article, we propose a novel collaborative learning method called SHIELD to defend against generative AF attacks. To expose AF signatures, we integrate an auxiliary generative model, called the defense (DF) generative model, which facilitates collaborative learning by combining input and output. Furthermore, we design a triplet model to capture correlations for real and AF attacked audios with real-generated and attacked-generated audios using auxiliary generative models. The proposed SHIELD strengthens the defense against generative AF attacks and achieves robust performance across various generative models. The proposed AF significantly reduces the average detection accuracy from 95.49\% to 59.77\% for ASVspoof2019, from 99.44\% to 38.45\% for In-the-Wild, and from 98.41\% to 51.18\% for HalfTruth for three different generative models. The proposed SHIELD mechanism is robust against AF attacks and achieves an average accuracy of 98.13\%, 98.58\%, and 99.57\% in match, and 98.78\%, 98.62\%, and 98.85\% in mismatch settings for the ASVspoof2019, In-the-Wild, and HalfTruth datasets, respectively.
\end{abstract}    
\section{Introduction}
\label{sec:intro}

\begin{figure}[!t]
    \centering
    \subfloat[\label{fig:intro_a}]{{\includegraphics[width=1\columnwidth]{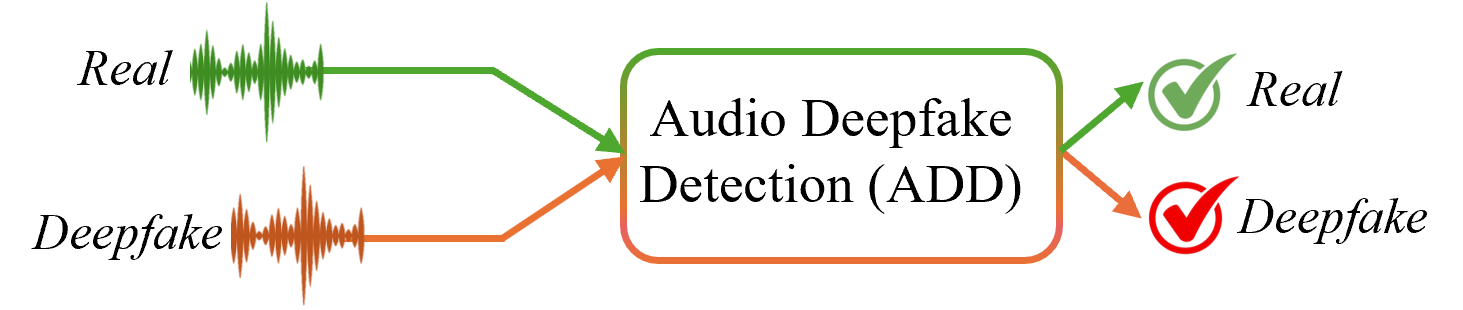}}} \\
    \subfloat[\label{fig:intro_b}]{{\includegraphics[width=1\columnwidth]{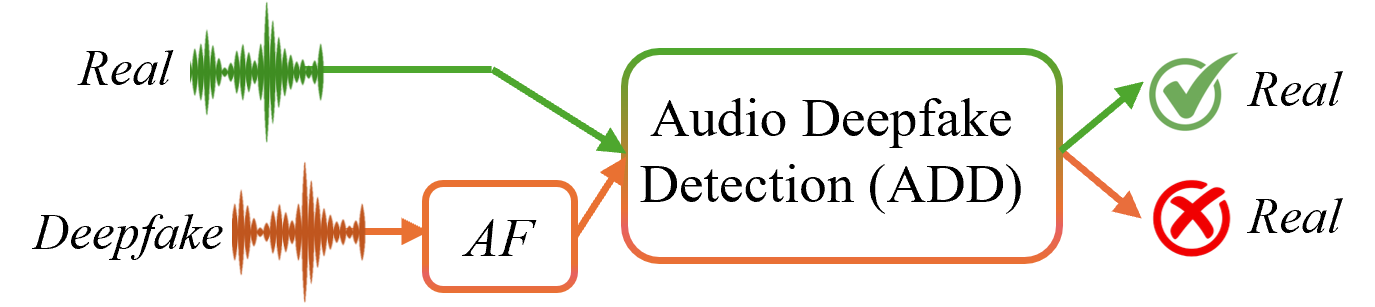}}} \\
    \subfloat[\label{fig:intro_c}]{{\includegraphics[width=1\columnwidth]{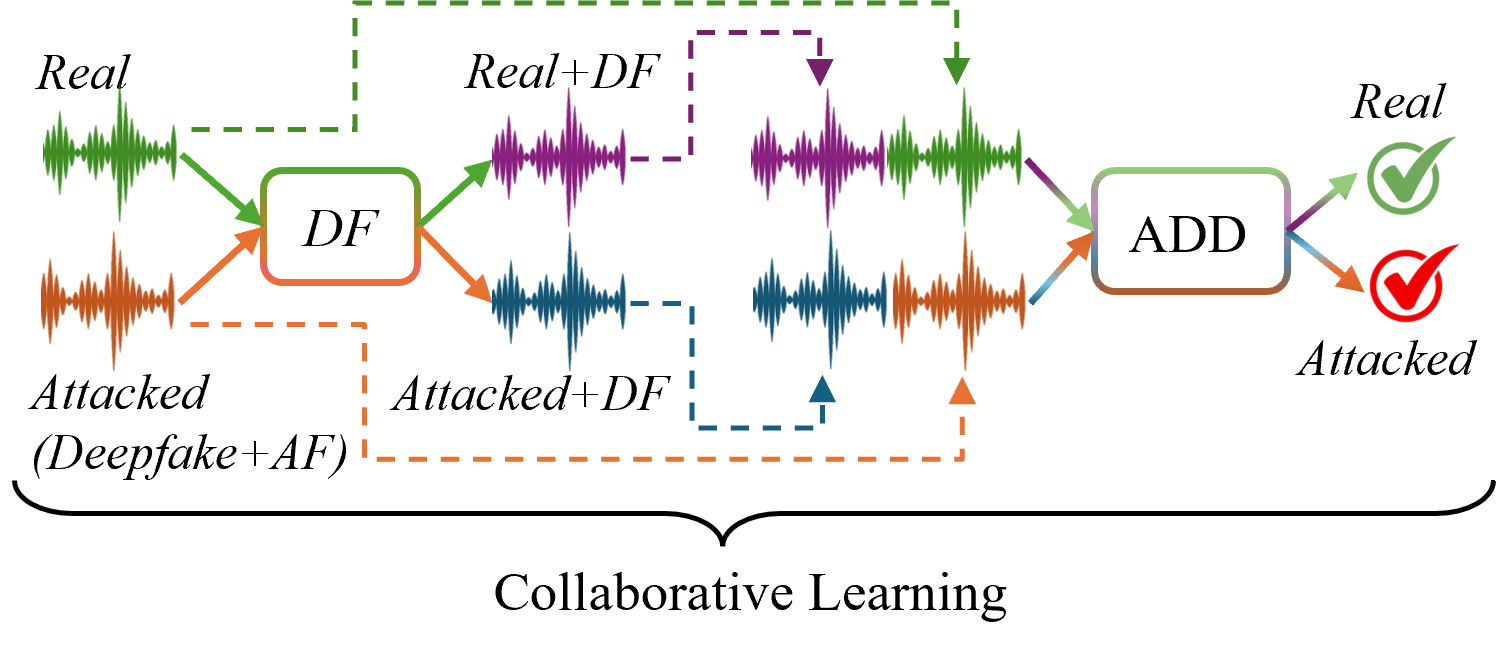}}}
    \caption{Conceptual overview of the traditional ADD, AF attacks, and proposed SHIELD framework against AF attacks.
    (a) Traditional ADD methods effectively distinguish between real and deepfake samples.
    (b) However, AF attacks applied to deepfakes reduce detection performance, making these methods vulnerable.
    (c) In contrast, the proposed collaborative learning-based SHIELD enhances robustness by incorporating auxiliary generative signatures (DF) to support the ADD model.}
    \label{fig:rel_fill}
\end{figure}
Advancements in generative artificial intelligence have significantly improved the generation of synthetic speech, enabling human-like audio deepfakes. Modern text-to-speech and voice cloning systems are now becoming capable of producing audio deepfake voices that sound almost the same as real human speech. This makes it difficult for human listeners, automatic speaker verification (ASV) systems, and deepfake detection tools to identify the difference between real and deepfake speech~\cite{khan2023battling,khan2023bridgingspoofgapunified}. These audio deepfakes have been exploited in serious threats such as identity fraud, misinformation campaigns, and high-impact social engineering attacks.

The threat is further amplified due to the accessibility of free or inexpensive voice cloning tools, which require minimal technical expertise to perform impersonation. These tools typically need short speech samples of a few seconds of a person's voice to copy their tone, pitch, accent, and speaking style. A recent report~\cite{realitydefender2024voice} shows that 37\% of the companies affected by identity fraud were targeted using AI-generated voice clones, making audio deepfakes the most common type of impersonation attacks. According to recent reports from IBM and the Identity Fraud Survey 2024~\cite{ibm2024deepfake}, deepfake-related fraud attempts have increased by 3,000\%, highlighting the urgent need for effective and reliable audio deepfake detection (ADD) methods.

To address this growing threat, the research community has developed various ADD systems~\cite{tak2021end,hua2021towards,khan2023spotnet,farooq2024securing,khan2023securing,farooq2025lightweight,khan2024frame} that rely on deep learning and feature representations to identify synthetic audio artifacts, as shown in Figure~\ref{fig:intro_a}. Although most deep learning-based ADD approaches show strong performance on known datasets and generation methods, they often fail to generalize to unseen data. More critically, most ADD are highly vulnerable to anti-forensic (AF) attacks, which introduce small, often imperceptible, perturbations to deception detection models as shown in Figure~\ref{fig:intro_b}. These attacks not only compromise the integrity of ADD methods but also raise concerns about their practical deployment in real-world applications to counter audio deepfakes. 

Although several defense strategies have been explored in the domains of image~\cite{uddin2023robust} and video~\cite{uddin2023deep} forensics, research on defending ADD methods against AF attacks, particularly generative AF attacks remains limited. Consequently, the existing literature lacks a crucial investigation into how AF attacks would affect ADD systems. Some notable studies~\cite{kawa2022defense, wu2020defense} have begun to address this gap. For example, Kawa et al.~\cite{kawa2022defense} and Wu et al.~\cite{wu2020defense} examined ASV-based systems under AF conditions and introduced spatial smoothing and adversarial training as potential defenses. Although these efforts address perturbation-based attacks, they overlook generative AF attacks.

More specifically, the existing work~\cite{kawa2022defense, wu2020defense} focuses mainly on traditional perturbation-based AF attacks on deepfakes and does not consider generative AF attacks generated via the generative adversarial network (GAN). Consequently, the impact of generative AF attacks on ADD methods remains an open and underexplored research area. To the best of our knowledge, this work is the first to propose a defense mechanism designed to protect against transferable generative AF attacks in ADD methods.

In this paper, we introduce a novel collaborative learning framework called SHIELD to enhance the robustness of ADD methods against generative AF attacks. Unlike traditional ADD pipelines as depicted in Figure~\ref{fig:intro_a}, the proposed SHIELD incorporates a defense (DF) generative model before the ADD stage, which facilitates collaborative learning between input and output representations, as illustrated in Figure~\ref{fig:intro_c}. An auxiliary DF generative model exposes AF artifacts by reconstructing inputs, revealing adversarial signatures. Additionally, we design a triplet-based model to capture intra- and inter-dependencies among real and AF attacked audio samples. This structure improves the model's ability to discriminate between real and deepfake inputs under adversarial conditions.

As illustrated in Figure~\ref{fig:intro_c}, conventional methods that rely solely on the traditional ADD mechanism in Figure~\ref{fig:intro_a} can be deceived by AF inputs, whereas the proposed method in Figure~\ref{fig:intro_c}, enhanced through collaborative learning, effectively mitigates these threats. The main contributions of this paper are as follows.

\begin{itemize}
    \item We introduce generative AF attacks on ADD methods and evaluate the vulnerabilities. 
    \item We propose a novel defense mechanism using a collaborative learning technique that improves the robustness of ADD methods against generative AF attacks.
    \item We apply a defense generative and triplet models that facilitate collaborative learning against AF signatures.
    \item We conducted comprehensive evaluations on deepfake benchmark datasets and demonstrated that AF attacks significantly degrade the performance of existing ADD methods. In contrast, the proposed defense method outperforms existing approaches, particularly in mitigating challenges posed by generative AF attacks.
\end{itemize}

 The remainder of this paper is organized as follows: Section 2 provides the existing literature on ADD methods and their vulnerabilities to AF attacks. Section 3 details the proposed collaborative learning framework, including the integration of the defense generative model and the triplet-based model to capture audio correlations. Section 4 describes the experimental setup, the datasets, evaluation metrics, and presents the results and discusses the effectiveness of the proposed approach compared to SoTA methods. Finally, Section 5 concludes the paper and suggests directions for future research.
 
\section{Related Works}
This section provides an overview of existing ADD methods, highlights progress in identifying their vulnerabilities to AF attacks, and discusses defense mechanisms proposed to address these challenges.

\subsection{Audio Deepfake Detection}
ADD methods generally fall into two categories: traditional methods based on handcrafted features (e.g., MFCCs, CQCCs, LFCCs) coupled with classifiers like Gaussian mixture models, and modern approaches leveraging deep learning to learn rich feature representations directly from raw or transformed audio for classification.

Deep learning approaches have gained prominence due to their ability to model high-dimensional audio representations. Consequently, in recent years, most ADD systems employed deep learning feature extraction and classification. For example, Tak et al.~\cite{tak2021end} introduced RawNet2, which processes raw waveforms through CNNs and GRUs to achieve state-of-the-art performance on ASVspoof2019. Although handcrafted feature extraction is eliminated, this approach exhibits sensitivity to unseen spoofing techniques. Similarly, Khan et al.~\cite{khan2023spotnet} introduced SpotNet, a CNN model that uses Mel-spectrograms optimized for edge devices. Extending this, Khan et al.~\cite{khan2024frame} later incorporated attention mechanisms to identify frame-level inconsistencies in synthetic speech, achieving robust detection on recent partial deepfake datasets such as HalfTruth.

Further advancements focus on leveraging pre-trained models for improving generalization and robustness. For instance, Zhang et al.~\cite{Zhang2024} used pre-trained rawboost and wav2vec-XLS-R for ADD, which reduces dependence on labeled data and improves generalization. Concurrently, Grinberg et al.~\cite{grinberg2025does} and Lim et al.,~\cite{lim2022detecting} integrated explainability techniques such as saliency maps and Grad-CAM to support forensic analysis. Wu et al.~\cite{wu2024clad} addressed non-adversarial manipulations through CLAD, a contrastive learning framework that minimizes sensitivity to volume changes and noise injection. Whereas, to preserve content privacy, Li et al. proposed SafeEar~\cite{li2024safeear}, a framework that decouples semantic and acoustic information, using only acoustic cues for detection.

Complementing these efforts, hybrid models with multi-feature fusion that fuse handcrafted and learned features have shown promising results in generalization. For instance, Yang et al.~\cite{yang2024robust} proposed a system that integrates handcrafted and learning-based features, achieving better generalization across datasets. Similarly, Ahmadiadli et al.~\cite{10337724} introduced an identity-independent detection system by combining handcrafted and neural features while explicitly modeling forgery-specific artifacts rather than speaker characteristics, which proved effective against previously unseen spoofing attacks.

Despite recent progress, many ADD systems remain tightly coupled to specific datasets and fail under real-world or adversarial conditions. This highlights the need for models with stronger generalization and robustness. Our work addresses this gap by introducing a detection framework that integrates transferable adversarial signatures, enabling more resilient and adaptive ADD.

\subsection{Vulnerabilities of Audio Deepfake Detection}
Although ADD systems have improved considerably, they remain vulnerable to adversarial attacks, which introduce imperceptible perturbations (often below -30 dB SNR) that manipulate decision boundaries while preserving audio fidelity for human listeners~\cite{rabhi2024audio}. These attacks expose key vulnerabilities in generalization, robustness to signal degradation, and resistance to transferable or generative attacks.

Initial vulnerability studies focused on gradient-based white-box attacks. For example, Kawa et al.~\cite{kawa2022defense} conducted a comprehensive evaluation of ADD systems, including RawNet3~\cite{tak2021end}, against adversarial attacks in both white-box and transferability scenarios. The results indicate that adversarial training, particularly adaptive training, can enhance model resilience. However, the study~\cite{kawa2022defense} mainly focused on traditional perturbation-based attacks and did not address more advanced generative adversarial strategies.

The threat landscape evolved with Farooq et al.~\cite{umar2025transferable} introduction of a transferable GAN-based adversarial attack framework. This approach successfully targeted different ADD systems, significantly reducing their accuracy in diverse scenarios and highlighting critical vulnerabilities in existing defenses. The presented ensemble-based method achieved 89\% attack success rates by exploiting latent space discontinuities, even across robust detectors, demonstrating that adversarial transfer learning can exploit shared vulnerabilities in ensemble models.

Following this, the emergence of generative adversarial networks has fundamentally transformed the threat landscape. For instance, Rabhi et al.~\cite{rabhi2024audio} demonstrated that even advanced detectors could be effectively bypassed using GAN-based adversarial attacks, reducing detection accuracy to nearly zero. Their findings emphasized the urgent need for generalizable, lightweight defense mechanisms for recent adversarial attacks. On the other hand, in recent investigations, frequency-domain vulnerabilities have been identified as particularly exploitable for ADD systems. For instance, Zhang et al.~\cite{zhang2024can} present F-SAT, a frequency-selective adversarial training method that focuses on high-frequency components, which are often exploited by attackers. The reported results demonstrated that the presented method improved detection accuracy on both clean and adversarial samples, thus highlighting the importance of frequency-domain characteristics in adversarial training.


Overall, these studies reveal a central limitation: most ADD systems are not inherently designed to overcome adversarial interference or to generalize beyond their training distributions. This shows the need for detection frameworks that can integrate adversarial robustness into their core architecture. The proposed approach addresses this gap by introducing transferable adversarial signatures directly into the detection pipeline, enabling more resilient and adaptive ADD.
\section{Method}
\label{sec:method}
\begin{figure}[!t]
    \centering
    \subfloat[\label{fig:arch_a}]{{\includegraphics[width=0.8\columnwidth]{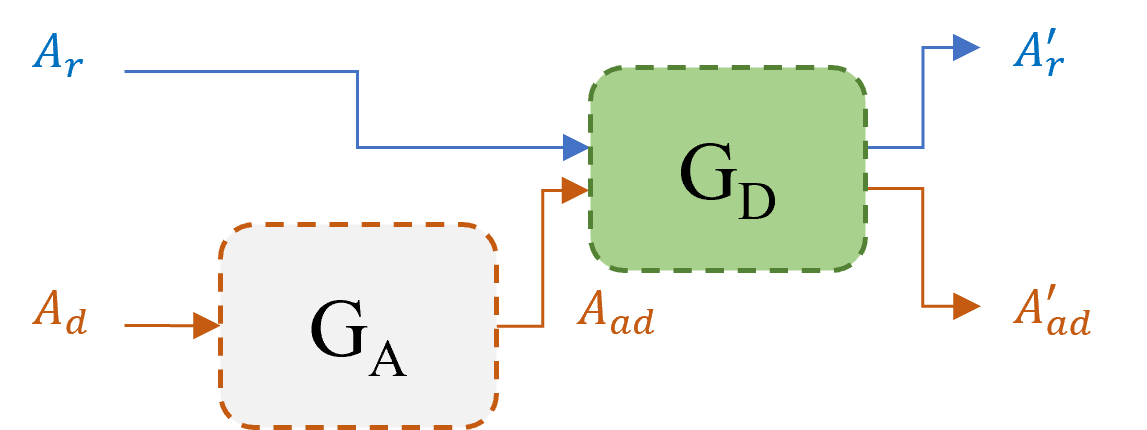}}}\\
    \subfloat[\label{fig:arch_b}]{{\includegraphics[width=1\columnwidth]{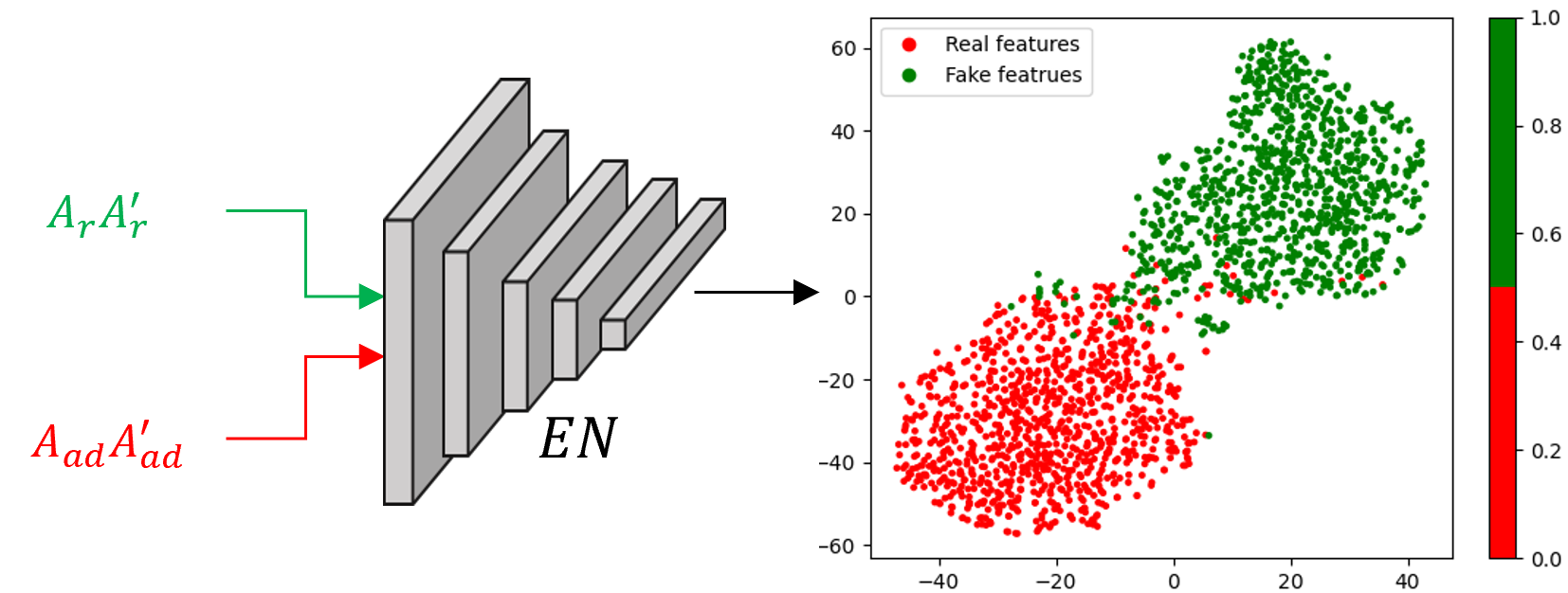}}}\\
    \caption{Proposed SHIELD architecture: (a) The deepfake audio ($A_d$)  is processed through an AF ($G_A$) model to generate AF attacked deepfake audio ($A_{ad}$). Subsequently, a defense ($G_D$) model is employed to generate real-generated ($A^{'}_r$) and attacked-generated ($A^{'}_{ad}$) audios. (b) The real and real-generated audios ($A_rA^{'}_r$), as well as the attacked and attacked-generated audios ($A_{ad}A^{'}_{ad}$), are concatenated to capture discriminative features using triplet learning. The integration of input and output to and from $G_D$ helps to explore the relationship among them to detect generative AF signatures.}
    \label{fig:arch}
\end{figure}

This section outlines the architectural overview, including collaborative learning, the training strategy of the GAN, and the triplet learning approach of the proposed defense.

\subsection{Problem Statement}
The rapid advancement of generative AI, particularly GANs, poses a significant challenge to ADD methods by creating more realistic deepfake audio. Alongside, GAN is widely used as an AF attack to alter the deepfake content without leaving visual signatures and deceive the ADD methods. Therefore, detecting generative AF attacks, particularly those performed by generative models, has become increasingly critical. To address this issue, we propose SHIELD, based on collaborative learning, designed to defend against generative AF attacks and secure ADD methods to detect AF attacked audio. We integrated a DF generative model to facilitate collaborative learning and provide robust audio attack detection. As generative AF leaves a distinct signature, adding a DF generative model before the ADD method helps explore the correlation between real vs. real-DF samples and attacked vs attacked-samples, enabling better separation. The hypothesis behind incorporating the DF generative model is that if a sample is real and passed through DF, the correlation between the original real and the real-DF versions will be low, as they exhibit different signatures. However, if a sample is AF-attacked and then passed through DF, the correlation between the attacked and attacked-DF versions will be high, since they share similar signatures.

\subsection{Architecture of the Proposed Method}
Figure~\ref{fig:arch} illustrates the workflow and architecture of the proposed method, which comprises two primary phases: (a) integration of a DF generative model, depicted at the top of Figure~\ref{fig:arch_a}, and (b) collaborative learning using a triplet model, shown at the bottom of Figure~\ref{fig:arch_b}. The details are presented in the following subsections. The AF attack is applied using a generative model ($G_A$) to generate AF-attacked deepfakes ($A_{ad}$) to deceive the ADD methods. In the proposed method, we integrate an auxiliary generative model ($G_D$) as a defense model that helps in collaborative learning. In the second phase, we use the combined samples to generate different embeddings for real and deepfake samples using triplet learning.

\subsubsection{Auxiliary Generative Model for Collaborative Learning}
Various AF techniques can be employed on deepfake audio to deceive ADD methods, ranging from traditional approaches ~\cite{wu2024clad} to generative ~\cite{umar2025transferable, uddin2019anti, uddin2021analysis} AF attacks. In this work, we focus specifically on detecting generative ~\cite{umar2025transferable} AF attacks, as it is more complicated and challenging to detect.

Assume $A_r$ and $A_d$ represent real and \{fake audio, respectively. In general, when an AF attack is applied using a generative model ($G_A$), it generates an attacked audio ($A_{ad}$), which is defined as follows:

\begin{equation}
A_{ad} = G_A(A_{d}) 
\end{equation}
When $A_{ad}$ is provided to the ADD method, it fails to correctly identify it as a deepfake and instead misclassifies it as real.

To defend against AF, we use an auxiliary generative model that works as a defense ($G_D$) to generate real-generated ($A^{'}_r$) and attacked-generated ($A^{'}_{ad}$) audios, defined as follows:

\begin{equation}
A^{'}_{r} = G_D(A_{r}) 
\end{equation}

and 
\begin{equation}
A^{'}_{ad} = G_D(A_{ad}) 
\end{equation}

We integrate $A_rA^{'}_r$ and $A_{ad}A^{'}_{ad}$ to learn collaboratively to distinguish between real and AF attacked samples. Here, $A_rA^{'}_r$ is considered as real, and $A_{ad}A^{'}_{ad}$ is considered as AF attacked samples. As $A_r$ and $A^{'}_r$ have the different signatures, for example, $A_r$ contains real information while $A^{'}_r$ contains AF signatures ($G_D$ signature), the correlation between $A_r$ and $A^{'}_r$ is very low. In contrast, $A_{ad}$ and $A^{'}_{ad}$ have the same generative signatures, for example, $A_{ad}$ contains $G_A$ signature while $A^{'}_{ad}$ contains $G_D$ signatures (both have the similar generative signatures), the correlation between $A_{ad}$ and $A^{'}_{ad}$ is very high. This allows high confidence to distinguish between real and AF attacked samples using deep triplet learning collaboratively.

\subsubsection{Training Strategy of GAN Model}
Since the focus is on performing generative AF attacks on deepfake audio to deceive SoTA ADD methods, we train the GAN model using a min-max optimization technique. In this setup, the generator's goal is to produce attacked deepfake audio that successfully evades detection without significantly distorting its quality. 

\begin{figure}[!t]%
\centering
{{\includegraphics[width=1\linewidth]{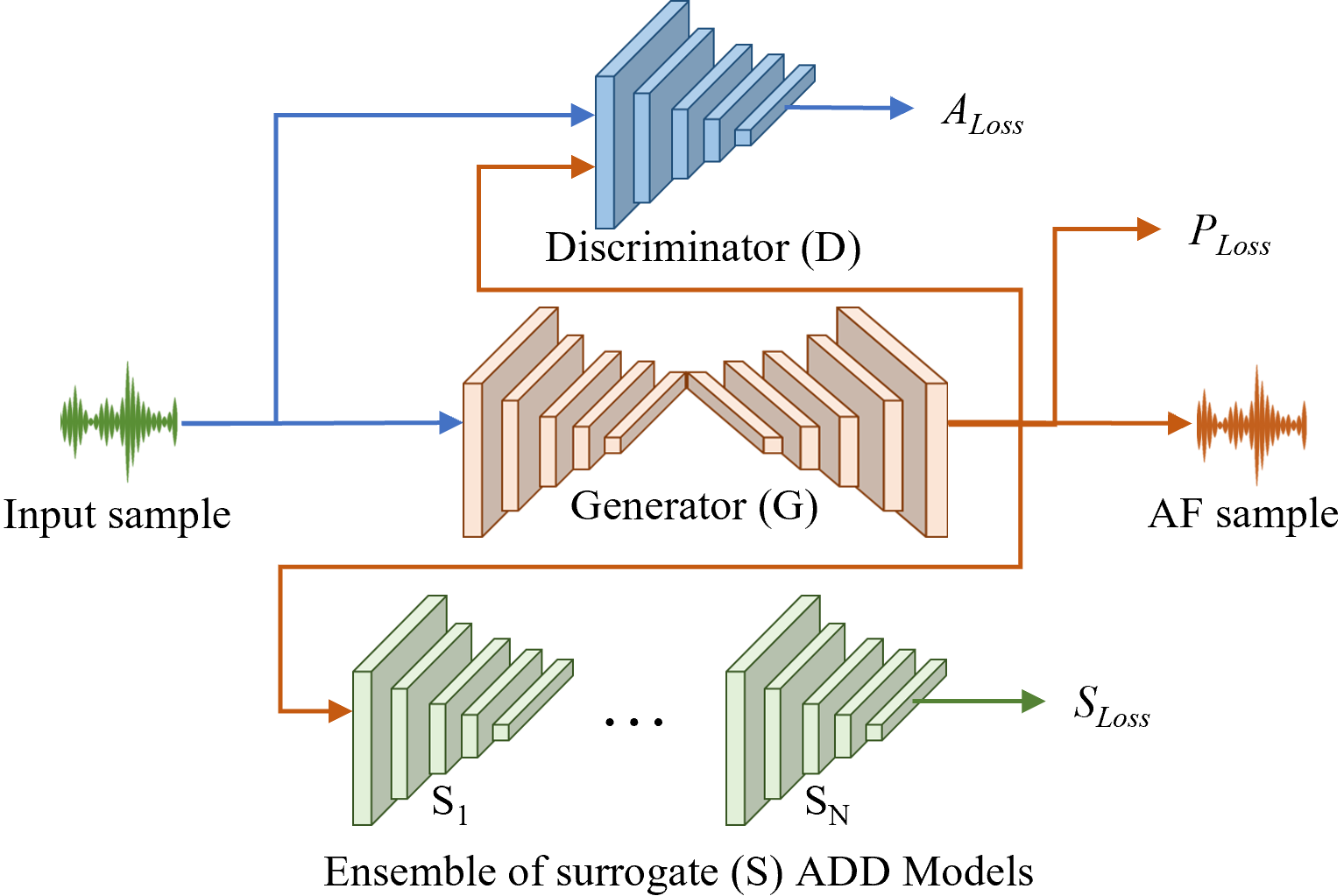}}}
    \caption{Training strategy of GAN-based models for generating AF attacks targeting ADD methods. The generator (G) network first applies AF to the input sample, producing an AF sample by receiving adversarial feedback from both the discriminator (D) and surrogate (S) models to deceive them.}
    \label{fig:af}
\end{figure}

To achieve this, we employ a combination of perceptual, adversarial, and surrogate losses during the training process ~\cite{umar2025transferable, uddin28enhanced} of the generator model as shown in Figure~\ref{fig:af}. The generator model receives feedback from both the discriminator and an ensemble of surrogate ADD methods, and updates its parameters accordingly to perform the AF attack.

We optimize the generator network with the combined loss function, defined as follows:

\begin{equation}
G_{loss} = P_{loss} + A_{loss} + S_{loss}
\end{equation}
where, $G_{loss}$ is the generator training loss. $P_{loss}$, $A_{loss}$, and $S_{loss}$ are the point-wise $L_1$, binary cross-entropy, and cross-entropy losses, respectively.

The $P_{loss}$, $A_{loss}$, and $S_{loss}$ are defined as follows:

\begin{equation}
P_{Loss} = \frac{1}{L} \sum_{i \in L}|(A_d[i] - {A}^{'}_d[i])|
\end{equation}

\begin{equation}
A_{Loss} = \log(1-D(G({A}^{'}_d)))
\end{equation}
and 

\begin{equation}
S_{Loss} = -\sum_{i\epsilon S}\sum_{j\epsilon L} \log(S_i(G({A}^{'}_d[i])))
\end{equation}  
where $L$ and $S$ are the length of the audio signal and the number of surrogate ADD methods.

The discriminator model is trained using a loss function to distinguish between real and attacked deepfake audio. This loss function is defined as follows:

\begin{equation}
D_{loss} = \log(1-D(A_r)) + \log(1-D(G(A_{d})))
\end{equation}
where $D_{loss}$ is the discriminator binary cross-entropy loss for real and deepfake samples. The first and second parts indicate the discriminator's real and deepfake losses, respectively, to apply the AF attack.

\subsubsection{Triplet Model for Generating Embedding}
After employing $G_D$, we concatenated $A_r$ and $A^{'}_r$ to represent real audio as $A_rA^{'}_r$. Similarly, $A_{ad}$ and $A^{'}_{ad}$ were concatenated to represent attacked audio as $A_{ad}A^{'}_{ad}$. Subsequently, we employed the triplet model ($TM$)~\cite{hoffer2015deep} to capture intra-class consistencies and inter-class inconsistencies, thereby generating discriminative features for effective detection. 

Let $F_a$, $F_p$, and $F_n$ represent embeddings generated by $TM$ for anchor ($a$), positive ($p$), and negative ($n$) samples, respectively. Then pairwise euclidean distance $D_{a\rightarrow p}$ between $a$ and $p$ and $D_{a\rightarrow n}$ between $a$ and $n$ are computed, defined as follows:

\begin{equation} 
D_{a\rightarrow p} = ||F_a - F_p||^2 
\end{equation}

and 

\begin{equation} 
D_{a\rightarrow n} = ||F_a - F_n||^2 
\end{equation}

We use margin ranking loss ($L_{mr}$) to train the triplet model, defined as follows:

\begin{equation} 
L_{mr} = \max(0, y \times (D_{a \rightarrow p} - D_{a \rightarrow n}) + m) 
\end{equation}
where $m$ is the margin value. We set it to 0.

After training the triplet model, we designed a simple fully-connected model to detect real and attacked audio.

\section{Experiments}
\label{sec:results}
\begin{table*}[t]
\centering
\caption{Performance of the baseline and GAN attacks for SoTA ADD methods.}
\begin{adjustbox}{width=1\textwidth}
\begin{tabular}{ccccccccccccc}
\toprule
\multicolumn{1}{c}{\multirow{2}{*}{\textbf{Methods}}} & \multicolumn{4}{c}{\textbf{ASVspoof2019}} & \multicolumn{4}{c}{\textbf{In-the-Wild}} & \multicolumn{4}{c}{\textbf{HalfTruth}} \\
\cline{2-13}
\multicolumn{1}{c}{} & \multicolumn{1}{c}{Baseline} & $G_1$ & $G_2$ & $G_3$ & \multicolumn{1}{c}{Baseline} & $G_1$ & $G_2$ & $G_3$ & \multicolumn{1}{c}{Baseline} & $G_1$ & $G_2$ & $G_3$ \\
\midrule
RawNet3~\cite{jung2022pushing} & 99.92 & 0.32 & 99.21 & 94.54 & 99.43 & 3.72 & 22.57 & 15.71 & 99.46 & 51.22 & 28.58 & 4.50 \\
RawNet2~\cite{tak2021end} & 96.00 & 77.63 & 86.14 & 89.68 & 99.48 & 0.77 & 3.31 & 20.98 & 98.76 & 71.03 & 69.79 &  70.12 \\
RawBoost~\cite{tak2021rawboost} & 90.68 & 56.15 & 72.90 & 89.62 & 99.79 & 2.00 & 4.40 & 51.32 & 97.63 & 69.27 & 85.04 &  97.63\\
Res-TSSDNet~\cite{hua2021towards} & 93.42 & 63.32 & 0.77 & 0.74 & 99.70 & 58.61 & 43.12 & 87.26 & 99.02 & 70.07 & 24.62 &  1.94\\
Inc-TSSDNet~\cite{hua2021towards} & 98.22 & 75.00 & 70.71 & 3.22 & 99.11 & 56.82 & 85.75 & 80.68 & 98.51 & 88.77 & 51.97 &  2.26\\
ResNet~\cite{hong2020holmes} & 97.82 & 63.54 & 86.94 & 0.25 & 99.57 & 63.20 & 16.31 & 0.38 & 99.30 & 99.89 & 17.48 &  0.02\\
MS-ResNet~\cite{wang2018csi} & 92.38 & 82.94 & 75.88 & 65.65 & 99.02 & 57.40 & 74.76 & 58.40 & 96.21 & 86.99 & 66.19 &  17.43\\
\midrule
\textbf{Avg.} & \textbf{95.49} & \textbf{59.84} & \textbf{70.36} & \textbf{49.10} & \textbf{99.44} & \textbf{34.65} & \textbf{35.75} & \textbf{44.95} & \textbf{98.41} & \textbf{76.75} & \textbf{49.10} & \textbf{27.70} \\
\bottomrule
\label{base_performance}
\end{tabular}
\end{adjustbox}
\end{table*}

\begin{figure*}[t!]
    \centering
    {{\includegraphics[width=1\linewidth]{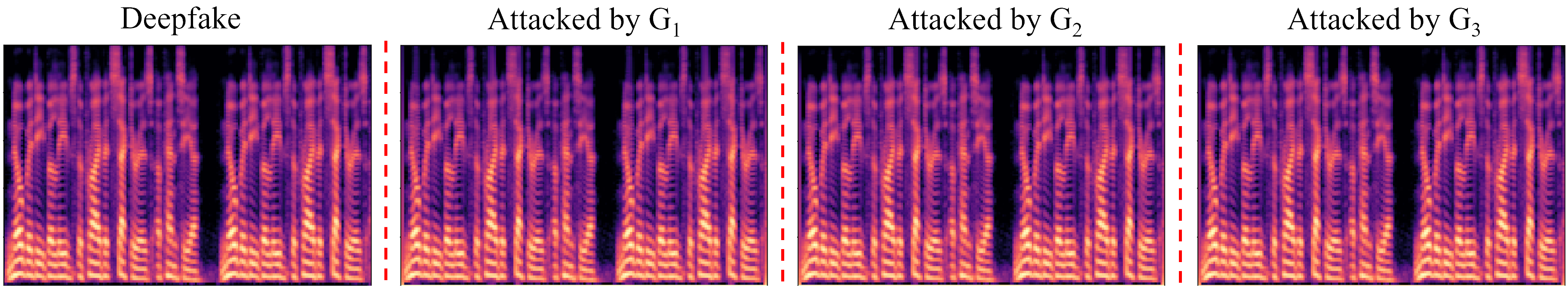}}} 
    \caption{Spectrogram visualization of deepfake and corresponding attacked samples generated by $G_1$, $G_2$ and $G_3$, respectively.} 
    \label{fi:vis_spec}
\end{figure*}

In this section, we describe the experimental results, including the datasets, the performance of the baseline ADD models, the application of generative AF attacks, and the proposed SHIELD mechanism to defend against AF attacks.

\subsection{Datasets}
To evaluate the performance of the baseline, AF, and proposed defense, we utilize three benchmark datasets: ASVspoof2019~\cite{todisco2019asvspoof}, HalfTruth~\cite{yi2021half}, and In-the-Wild~\cite{muller2022does}. These datasets provide diverse real and deepfake samples generated by various speech synthesis and voice conversion techniques. \\
\textbf{ASVspoof 2019~\cite{todisco2019asvspoof}}: The ASVspoof 2019~\cite{todisco2019asvspoof} dataset is a widely used benchmark for assessing vulnerabilities in ASV systems. It includes logical and physical access scenarios, with the logical subset containing 121,461 audio samples generated using 17 different TTS and VC algorithms. Further details about the parametric configuration can be found in ~\cite{todisco2019asvspoof}. \\
\textbf{HalfTruth~\cite{yi2021half}}: The HalfTruth dataset~\cite{yi2021half} is designed to assess the detection of partial audio deepfakes, containing 53,612 samples that blend synthetic speech from advanced TTS and voice cloning technologies with genuine recordings. It features contributions from 5,592 male and 20,962 female voices, ensuring diverse representation. Further details can be found in~\cite{yi2021half}.\\
\textbf{In-the-Wild~\cite{muller2022does}}: The In-the-Wild dataset contains real and synthetic audio collected from interviews, online media, and other real-world sources, incorporating noise and environmental variations. Unlike ASVspoof2019~\cite{todisco2019asvspoof} and HalfTruth~\cite{yi2021half}, which contain controlled synthetic speech, In-the-Wild~\cite{muller2022does} presents deepfake scenarios in unconstrained conditions. 

\subsection{Experimental Setup}
All experiments were conducted on a high-performance Lambda server equipped with 48GB of GPU memory and 180GB of RAM, ensuring efficient processing of large-scale audio datasets. We utilized PyTorch as the primary deep learning framework, leveraging CUDA 12.1 for accelerated computations on NVIDIA RTX 6000 ADA GPUs.  

We trained the SoTA ADD methods using a cross-entropy loss function for 50 epochs, with a batch size of 256 and a learning rate of 0.0001. We train the GAN models for 30 epochs, which yields the best performance. The batch size and learning rate are set to 32 and 0.0001, respectively, for both the generator and the discriminator. Additionally, we use pretrained ADD methods as an ensemble of surrogate models. We train the triplet model for 50 epochs to achieve the best performance. The batch size and learning rate are set to 32 and 0.0001, respectively. We applied mixed-precision training for optimized training. We optimize all the models using the Adam optimizer. To improve the generalization on unseen test sets, we combined the training data from all three datasets to train all models.

\begin{table}[!b]
  \centering 
  \caption{Evaluation of the proposed defense mechanism in the match setting.}
  \begin{adjustbox}{width=0.5\textwidth}
  \begin{tabular}{lcccc}
    \toprule
    \textbf{Setting}  & \textbf{ASVspoof2019}  & \textbf{In-the-Wild} & \textbf{HalfTruth} \\
    \midrule
    $G_1\rightarrow G_1$ & 97.42 & 98.75 & 99.32  \\
    $G_2\rightarrow G_2$ & 98.85 & 99.04 & 99.61  \\
    $G_3\rightarrow G_3$ & 98.12 & 97.95 & 99.77  \\
    \midrule
    \textbf{Average}  & \textbf{98.13}  & \textbf{98.58}   & \textbf{99.57} \\
    \bottomrule 
  \end{tabular}
  \label{tab_defense_same_gans}
  \end{adjustbox}
\end{table} 

\begin{table}[!b]
  \centering 
  \caption{Evaluation of the proposed defense mechanism in the mismatch setting.}
  \begin{adjustbox}{width=0.5\textwidth}
  \begin{tabular}{lccc}
    \toprule
    \textbf{Setting}  & \textbf{ASVspoof2019}  & \textbf{In-the-Wild} & \textbf{HalfTruth} \\
    \midrule
    $G_1\rightarrow G_2$ & 99.12 & 98.85 & 99.43 \\
    $G_1\rightarrow G_3$ & 99.75 & 99.04 & 99.61 \\
    \midrule
    \textbf{Avg. ($G_1$)} & \textbf{99.44} & \textbf{98.95} & \textbf{99.52} \\
    \midrule
    $G_2\rightarrow G_1$ & 96.74 & 98.32 & 98.56 \\
    $G_2\rightarrow G_3$ & 98.71 & 97.54 & 98.24 \\
    \midrule
    \textbf{Avg. ($G_2$)} & \textbf{97.73} & \textbf{97.93} & \textbf{98.40} \\
    \midrule
    $G_3\rightarrow G_1$ & 98.89 & 99.01 & 99.16 \\
    $G_3\rightarrow G_2$ & 99.44 & 98.96 & 99.08 \\
    \midrule
     \textbf{Avg. ($G_3$)} & \textbf{99.17} & \textbf{98.99} & \textbf{99.12} \\
     \midrule
    \textbf{Overall Avg.} & \textbf{98.78} & \textbf{98.62} & \textbf{98.85} \\
    \bottomrule 
  \end{tabular}
  \label{tab_defense_diff_gans}
  \end{adjustbox}
\end{table}

\begin{table}[!b]
  \centering 
  \caption{Performance comparison of the proposed method with SoTA defense mechanism.}
  \begin{adjustbox}{width=0.5\textwidth}
  \begin{tabular}{lcccc}
    \toprule
    \textbf{Setting}  & \textbf{ASVspoof2019}  & \textbf{In-the-Wild} & \textbf{HalfTruth} \\
    \midrule
     SENet~\cite{wu2020defense}, 2020 & 68.34 & 53.11 & 57.54  \\
     LCNN~\cite{kawa2022defense}, 2023 & 85.12 & 60.44 & 60.22  \\
     ABC-CapsNet~\cite{wani2024abc}, 2024 & 66.23 & 58.64 & 66.79  \\
    \midrule
    \textbf{Proposed}& \textbf{98.78} & \textbf{98.62} & \textbf{98.85} \\
    \bottomrule 
  \end{tabular}
  \label{tab_defense_com}
  \end{adjustbox}
\end{table} 

\subsection{Performance Evaluations and Comparisons}
This section provides a detailed evaluation of the baseline ADD methods, AF attacks, and the proposed SHIELD against AF attacks. 

\subsubsection{Baseline Deepfake Detection and Anti-Forensic Attack Results on Benchmark Datasets}
We selected seven SoTA ADD methods, including RawNet3\cite{jung2022pushing}, RawNet2\cite{tak2021end}, RawBoost\cite{tak2021rawboost}, Res-TSSDNet\cite{hua2021towards}, Inc-TSSDNet\cite{hua2021towards}, ResNet\cite{hong2020holmes}, and MS-ResNet~\cite{wang2018csi}, to evaluate both baseline performance and AF attacks on benchmark deepfake datasets. These methods represent the most advanced detection techniques that leverage both raw waveform and spectrogram-based analyses to expose audio unnaturalness. 

Table~\ref{base_performance} presents the baseline performance of ADD methods across three different datasets, such as ASVspoof2019, In-the-Wild, and HalfTruth. The results show that these ADD methods effectively differentiate real and deepfake audio samples, achieving an average accuracy of 95.49\% on ASVspoof2019, 99.44\% on In-the-Wild, and 98.41\% on HalfTruth, respectively. 

To conduct AF attacks on ADD methods, we explored SoTA GAN models for audio processing and fine-tuned three architecturally distinct models: $G_1$ (UNet)\cite{ebner2020audio}, $G_2$ (SEGAN)\cite{pascual2017segan}, and $G_3$ (OPGAN)\cite{ince2022blind}. These models were selected based on their diverse architectures and effectiveness in speech synthesis and enhancement, making them suitable for AF attacks. 

The attack results, presented in Table~\ref{base_performance}, reveal a significant decline in ADD detection performance. Specifically, the average accuracy dropped to 59.84\%, 70.36\%, and 49.10\% on ASVspoof2019; 34.65\%, 35.75\%, and 44.95\% on In-the-Wild; and 76.75\%, 49.10\%, and 27.70\% on HalfTruth when using $G_1$, $G_2$, and $G_3$, respectively. These results demonstrate the vulnerability of existing ADD methods to adversarial perturbations, highlighting the need for more robust countermeasures. 

\subsubsection{Performance Evaluation of the Proposed SHIELD against Generative AF Attacks}
We carried out two settings to evaluate the proposed SHIELD against generative AF attacks. First, match scenario, in which the attack ($G_A$) and defense ($G_D$) use the same generator model as shown in Figure~\ref{fig:arch}. Second, a mismatch scenario, where the attack ($G_A$) and defense ($G_D$) use different generator models. RawNet3~\cite{jung2022pushing} is employed as the embedding network in our triplet model to collaboratively learn discriminative features. Its selection is motivated by its superior performance compared to SoTA methods, as shown in Table~\ref{base_performance}.

The defense results in the match setting are presented in Table~\ref{tab_defense_same_gans} in which we use the same GAN model to apply AF attacks and defend against them such as $G_1\rightarrow G_1$, $G_2\rightarrow G_2$, and $G_3\rightarrow G_3$. The proposed method achieved average detection accuracies of 98.13\% for the ASVspoof2019, 98.58\% for the In-the-wild, and 99.57\% for the HalfTruth datasets, respectively. This confirms that the proposed SHIELD can effectively detect known generative AF attacks on deepfakes and offers a robust solution for ADD method.

Table~\ref{tab_defense_diff_gans} summarizes the detection performance of the proposed SHIELD mechanism in the mismatch setting. In the mismatch setting, we use different combination of three GAN models, such as $G_1\rightarrow G_2$, $G_1\rightarrow G_3$, $G_2\rightarrow G_1$, $G_2\rightarrow G_3$, $G_3\rightarrow G_1$, and $G_3\rightarrow G_2$ to evaluate the performance. Similar to the match setting, the method demonstrated strong effectiveness in countering unknown AF attacks on deepfakes. The proposed SHIELD achieved overall average accuracies of 98.78\% for ASVspoof2019, 98.62\% for In-the-Wild, and 98.85\% for HalfTruth datasets, respectively. Compared to baseline ADD methods, the proposed defense achieved superior performance in mitigating generative AF attacks in either setting.

\begin{figure}[!b]%
    \centering
{{\includegraphics[width=1\linewidth]{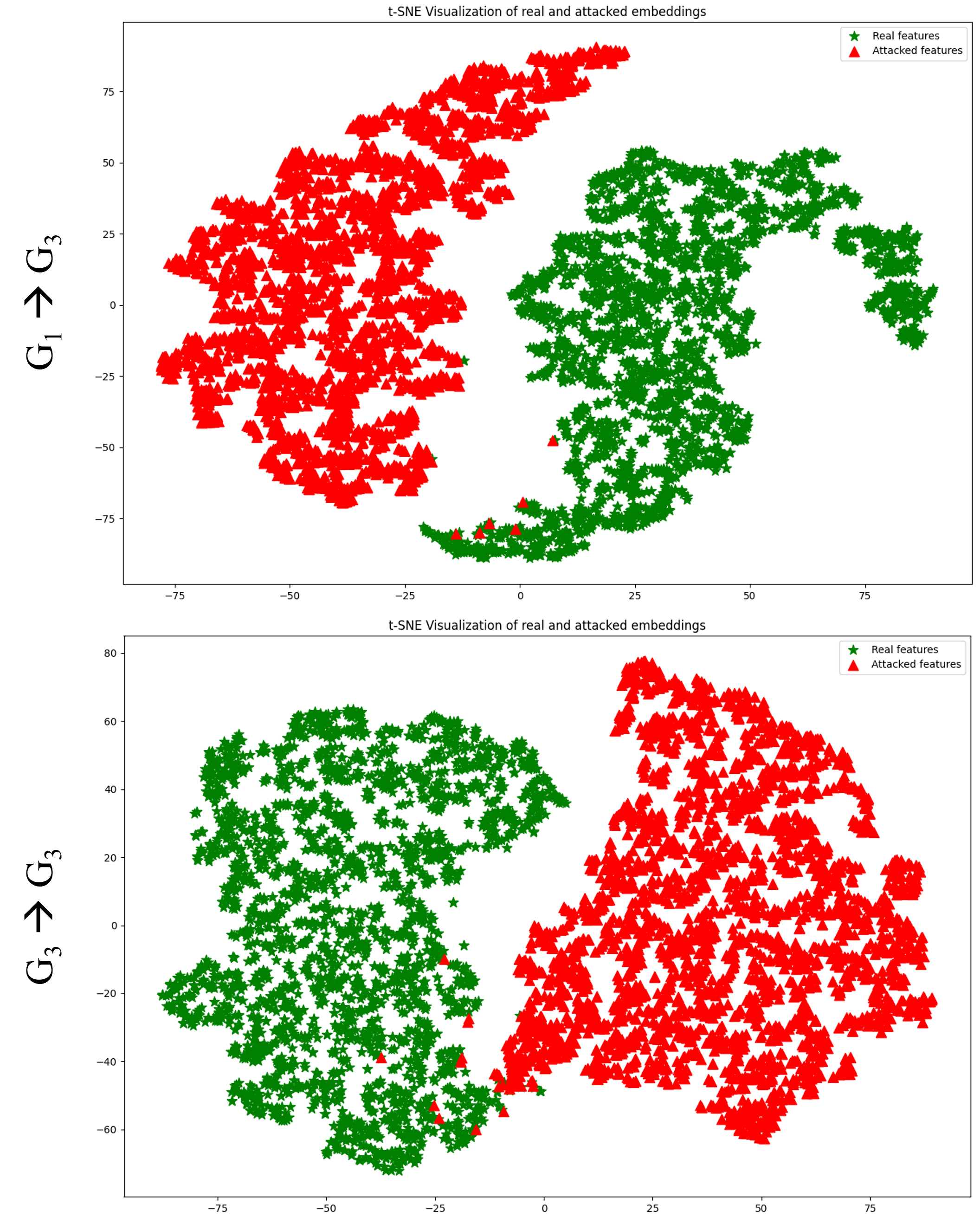}}}
    \caption{t-SNE visualization of embeddings for HalfTruth datasets produced by the triplet model.}
    \label{fi:vis_tsne}
\end{figure}

\subsubsection{Performance Comparisons}
We compared the proposed SHIELD with SoTA defense mechanisms, such as ABC‑CapsNet~\cite{wani2024abc}, LCNN~\cite{kawa2022defense}, and SENet~\cite{wu2020defense}. We computed the average robustness results of the SoTA defenses against $G_1$, $G_2$, and $G_3$, respectively. Table~\ref{tab_defense_com} compares the detection results of the proposed SHIELD with three SoTA methods. The proposed SHIELD outperforms existing defense methods by achieving improvements of 30.44\%, 45.51\%, and 41.31\% over SENet; 13.66\%, 38.18\%, and 38.63\% over LCNN; and 32.55\%, 39.98\%, and 32.06\% over ABC-CapsNet on the ASVspoof2019, In-the-Wild, and HalfTruth datasets, respectively, showing that \textsc{SHIELD} consistently surpasses these SoTA defenses under AF attacks.

\subsection{Ablation Study}
In our experiments and analysis, we employ three GAN models to execute AF attacks. To better understand, we visualize the spectral characteristics of both deepfake and AF attacked samples, as illustrated in Figure~\ref{fi:vis_spec}, to facilitate a comparative quality assessment. The visualization reveals that differentiating between deepfake and AF attacked samples poses significant challenges for human perception. Moreover, this AF attacks proficiently circumvent SoTA ADD methods, highlighting their efficacy and the inherent limitations of current detection techniques.

To illustrate the efficacy of the triplet model, we present t-SNE representations in Figure~\ref{fi:vis_tsne}, utilizing RawNet3~\cite{jung2022pushing} on the HalfTruth dataset. Specifically, we show results for the $G_3 \rightarrow G_3$ (bottom) match and $G_1 \rightarrow G_3$ (top) mismatch configurations. The visualization distinctly demonstrates that the triplet model effectively clusters genuine and AF attacked samples, rendering them easily distinguishable.
\section{Conclusion and Future Works}
\label{sec:conclusions}

The revolution of generative models has enabled the generation of more realistic deepfakes without leaving any visual clues to distinguish them from real samples. Therefore, several ADD methods have been proposed to detect generative signatures to reveal deepfakes.
Despite advancements in ADD, these methods remain vulnerable to generative AF attacks. To address this, we introduced a novel SHIELD mechanism that incorporates an auxiliary generative model as a defense within a collaborative learning framework. By leveraging triplet learning, our approach effectively captures generative signatures in deepfake audio, improving detection performance across benchmark datasets in both matched and mismatched scenarios.

While our method effectively counters generative AF attacks, future work will extend its applicability to a wider range of adversarial threats, such as diffusion, filtering, and other techniques like noise injection attacks, time-frequency manipulation, and temporal attacks. Additionally, integrating multi-modal analysis could enhance robustness, as adversarial techniques often target individual modalities like audio or visual data.
{
    \small
    \bibliographystyle{ieeenat_fullname}
    \bibliography{main}
}

\end{document}